\documentclass[12pt]{article}

\normalbaselines \topmargin -0.5 truein  \oddsidemargin 0.10
truein \evensidemargin 0.10 truein 
scaled \magstep2  \font\tbf = cmbx12

\begin{document}


\indent \vskip 1cm \centerline{\tbf  ANOMALOUS CHARACTER OF THE
AXION - PHOTON COUPLING}

\centerline{\tbf IN A MAGNETIC FIELD DISTORTED BY }

\centerline{\tbf A PP-WAVE GRAVITATIONAL BACKGROUND }

\vskip 0.8cm


 \vskip 0.3cm \centerline{\tbf Alexander B.
Balakin\footnote{e-mail: Alexander.Balakin@kpfu.ru} } \vskip 0.3cm
\centerline{\it Kazan Federal University, Kremlevskaya str. 18,
420008, Kazan,  Russia,} \vskip 0.5cm \centerline{and} \vskip
0.5cm \centerline{\tbf Wei-Tou Ni\footnote{e-mail:
weitou@gmail.com} } \vskip 0.3cm \centerline{\it Center for
Gravitation and Cosmology, Department of Physics,} \centerline{\it
National Tsing Hua University, Hsinchu, Taiwan 30013, Republic of
China} \vskip 3cm

{\tbf Abstract} \quad {\small We study the problem of axion-photon
coupling in the magnetic field influenced by gravitational
radiation. We focus on exact solutions to the equations for axion
electrodynamics in the pp-wave gravitational background for two
models with initially constant magnetic field. The first model
describes the response of an initially constant magnetic field in
a gravitational-wave vacuum with unit refraction index; the second
model is characterized by a non-unit refraction index prescribed
to the presence of ordinary and/or dark  matter. We show that both
models demonstrate anomalous behavior of the electromagnetic field
generated by the axion-photon coupling in the presence of magnetic
field, evolving in the gravitational wave background. The role of
axionic dark matter in the formation of the anomalous response of
this electrodynamic system is discussed.}



\newpage

\section{Introduction}

Axion-photon conversion in a strong magnetic field is nowadays
studied  by many experimental groups (see, e.g., the reports of
Collaborations abbreviated as PVLAS \cite{PVLAS,PVLAS2}, GammeV
\cite{GammeV}, CAST \cite{CAST,CAST2}, OSQAR \cite{OSQAR}, Q \& A
\cite{QA,QA2}, BMV \cite{BMV,BMV2}). These investigations are motivated by
the search for new light pseudo-bosons (axions)
\cite{PQ,Weinberg,Wilczek} forming (hypothetically) the dark
matter \cite{ADM1,ADM2,ADM3}, which is considered to be a key
element in the structure of our Universe. The main physical
mechanism, on which the corresponding experiments are based can be
described in the frameworks of axion electrodynamics
\cite{Ni,Sikivie,Wlczk}. According to this theory, the axions can
be created by the electromagnetic field, for which the electric
and magnetic components are not orthogonal one to another; the
simplest variant in this sense is to use the combination of static
magnetic field and electromagnetic wave propagating
perpendicularly to it, so that the oscillating electric field is
directed along the static magnetic one. This idea is the base,
e.g., for the Light Shining through the Wall (LSW) experiments. On
the other hand, if we are surrounded by dark matter axions, the
axion electrodynamics predicts the phenomenon of axionically
induced optical activity (see, e.g., \cite{Ni2}), if the
pseudoscalar field $\phi$, related to axions, has non-vanishing
gradient four-vector $\nabla_k \phi \neq 0$. According to a
standard classification the invariant $I {=} g^{ik} \nabla_i \phi
\nabla_k \phi$ can be positive (e.g., for cosmological model with
$\phi(t)$ depending on time only), negative (e.g., for static
spherically symmetric system with $\phi(r)$ depending on the
radial variable only), and can be equal to zero (e.g., for systems
with pp-wave symmetry, for which pseudoscalar field depends on the
retarded time only, $\phi(x{-}ct)$, so that the four-gradient
$\nabla_k \phi$ is the so-called null four-vector). In other
words, when the pseudo-scalar (axion) and magnetic fields are
constant, the axion-photon coupling is hidden, and some
non-stationary external field is necessary to activate it.

Our idea is to use a non-stationary gravitational field in order
to activate the frozen axion-photon coupling in the static
magnetic field in the axion dark matter environment. We suggest to
use for this purpose the field of gravitational radiation incoming
from periodic astrophysical sources. The amplitude of such
gravitational waves from these distant sources is weak, and
formally speaking it seems to be a pessimistic detail.
Nevertheless, we have found that the axionic dark matter
environment can (theoretically) play  not only the role of a
mediator, but also the role of an {\it amplifier} of the signal -
response in the process of gravitational wave action on the
magnetic field.

In the frameworks of pure electrodynamics the problem of
interaction of weak gravitational waves with static electric and
magnetic fields in vacuum was studied in seventies of the last
century (see, e.g., \cite{boc,zel1}). In the case, when
a pp-wave gravitational background distorts a magnetic field in
the dielectric environment with non-unit refraction index, $n^2
\neq 1$, the behavior of the corresponding electromagnetic
response, as was shown in \cite{BV,BL}, becomes critical, i.e., it
can be amplified anomalously, if $n^2 \to 1$.

In this paper we extend the theory of interaction between
gravitational and electromagnetic fields and consider an exactly
solvable model of evolution of an initially static and homogeneous
magnetic and pseudoscalar fields in the non-linear gravitational
wave background. This model is new, since one extra ingredient,
namely axion field, is added into the scheme of interaction
considered in \cite{BL}. We show that the discussed new mechanism
of the axion-photon-graviton coupling can produce anomalous
electric field response.

The paper is organized as follows. In Section 2 we describe the
model using the Lagrangian formalism, and derive (in general form) the equations of axion
electrodynamics in a dielectric medium (and vacuum), as well as evolutionary equations
for the gravitational field and for macroscopic velocity of the medium.
In Section 3 we reduce the equations obtained in Section 2 to the case, when the space-time
possesses the plane-wave symmetry: in Section 3.1 we discuss the properties of the
gravitational-wave background; in Section 3.2 we describe the initial state of the electrodynamic system,
i.e., the state before the gravitational wave appearance (constant magnetic field, constant axion field and vanishing electric field);
in Section 3.3 we rewrite the equations of axion electrodynamics
in the form coordinated with the chosen space-time symmetry.
Section 4 is devoted to the analysis of the model in the case, when the refraction index is equal to one (axionic vacuum):
in Section 4.1 we present the exact (and unique) solution to the equations of axion electrodynamics in the gravitational wave background for the case,
when the initial magnetic field is arbitrarily directed with respect to the front of the plane gravitational wave; in Section 4.2  we discuss
two special exact solutions.  In Section 4.3 we analyze the physical properties of the obtained exact solution: gravitationally induced distortions
of the initial magnetic field, the generation of an electric field in the axionic environment, the anomalous character of the obtained solutions.
In Section 5 we obtain the exact solution to the equations of axion electrodynamics in the framework of the model with non-unit
refraction index, and analyze the critical properties of this solution. In Section 5.3 we summarized the features of obtained exact solutions; in Section 5.4 we propose for discussion
our explanation of the critical behavior of the obtained solutions in terms of phase transition of the second kind.
Section 6 is devoted to applications of two
studied models to the possible experiments with magnetic field in the axionic background under the influence of the periodic gravitational radiation from relativistic binary.
In Section 6.1 we obtain working formulas for a weak gravitational-wave field based on the exact solutions discussed above. In Section 6.2 we discuss the estimations of the predicted effects
for the axionic vacuum (6.2.1) and for the medium with non-unit refraction index (6.2.2), and focus on the constraints of the model (6.2.3).
In Section 7 we discuss the described new mechanism of the
axion-photon-graviton coupling and estimations of the effect
magnitude for the terrestrial magnetic field and magnetized interstellar medium.

\section{The model}

\subsection{Action functional}

Let us start from the action functional
$$
S = \int d^4 x \sqrt{{-}g} \left\{\frac{R{+}2\Lambda}{2\kappa} + L_{({\rm matter})} +
\frac{1}{4} C^{ikmn}F_{ik} F_{mn} + \frac{1}{4} \phi F^{*mn}F_{mn}
+ \right.
$$
\begin{equation}
\left. + \frac12 \Psi^2_0 \left[ {-} g^{mn} \nabla_m \phi \nabla_n \phi
{+} m^2_{({\rm A})} (\phi^2 {-} \phi^2_*)  + \frac12 \lambda
\left(\phi^2-\phi^2_* \right)^2\right]\right\} \,. \label{actmin}
\end{equation}
Here $g$ is the determinant of the metric tensor $g_{ik}$,
$\nabla_{m}$ is a covariant derivative, $R$ is the Ricci scalar,
$\kappa = \frac{8 \pi G}{c^4}$ is the Einstein constant and
$\Lambda$ is the cosmological constant. As usual, $F_{mn}$ is the
Maxwell tensor, $F^{*mn} \equiv \frac{1}{2} \epsilon^{mnpq}F_{pq}$
is the tensor dual to $F_{pq}$, $\epsilon^{mnpq} \equiv
\frac{1}{\sqrt{-g}} E^{mnpq}$ is the Levi-Civita tensor,
$E^{mnpq}$ is the absolutely antisymmetric Levi-Civita symbol with
$E^{0123}=1$. The Maxwell tensor may be represented in terms of a
four-vector potential $A_i$ as
\begin{equation}
F_{ik} = \nabla_i A_{k} - \nabla_k A_{i} \,, \label{maxtensor}
\end{equation}
so the dual Maxwell tensor satisfies the condition
\begin{equation}
\nabla_{k} F^{*ik} =0 \,. \label{Emaxstar}
\end{equation}
The term $L_{({\rm matter})}$ describes Lagrangian of a matter;
it can depend on the potential four-vector $A_i$ itself but does not contain the Maxwell tensor.
The quantity $C^{ikmn}$ describes a linear response tensor; in
this work we use the following model representation of this
tensor:
\begin{equation}
C^{ikmn} {=} \frac{1}{2\mu}\left[(g^{im}g^{kn} {-} g^{in}g^{km})
{+} (n^2{-}1)\left( g^{im} U^kU^n {-} g^{in}U^kU^m {+}
g^{kn}U^iU^m {-} g^{km}U^iU^n \right)\right] \,. \label{Cikmn}
\end{equation}
Here $n$ is the refraction index defined as $n^2{}= \varepsilon
\mu$, where $\varepsilon$ and $\mu$ are the constants
characterizing dielectric and magnetic permittivities of the
medium. The term $U^i$ denotes the macroscopic velocity
four-vector of the medium; we assume that this four-vector is chosen to be the
time-like eigen-vector of the stress-energy tensor of the matter.
When $\varepsilon{=}1$, $\mu{=}1$ and
thus $n^2{=1}$, we obtain the model with pure vacuum, for which
the term $C^{ikmn}F_{ik}F_{mn}$ in the action functional
transforms into the standard term $F^{ik}F_{ik}$. Let us mention
that the term $\phi F^{*mn}F_{mn}$ also could be included into
$C^{ikmn} F_{ik}F_{mn}$ by extending the linear response tensor
$C^{ikmn} \to {\cal C}^{ikmn} = C^{ikmn}{+} \frac12 \phi \
\epsilon^{ikmn}$, nevertheless, we prefer to visualize it as a
specific term describing classical axion-photon coupling.

The symbol $\phi$ stands for a pseudo-scalar field, this quantity
is dimensionless providing the terms $\frac{1}{2} F^{mn}F_{mn}$
and $ \frac{1}{2} \phi F^{*mn}F_{mn}$ to have the same
dimensionality. The axion field itself, $\Phi$, is considered to
be proportional to this quantity $\Phi = \Psi_0 \phi$ with a
constant $\Psi_0$ related to the vacuum averaged value of this
field.  The term $m_{({\rm A})}$ is proportional to a
(hypothetical) mass of an axion, $m_{({\rm A})} = c \ m_{({\rm
axion})} / \hbar$; $\hbar$ is the Planck constant. The constant
$\phi_*$ relates to an averaged vacuum value of the axion field,
and $\lambda$ is a coupling constant of the fourth-order
self-interaction of the axion field. Formally speaking, the term
with the cosmological constant $\frac{2\Lambda}{\kappa}$ can
absorb the constant ${-}\Psi^2_0 m^2_{({\rm A})} \phi^2_*$,
nevertheless, we prefer this form of decomposition of the
potential. Moreover, we consider below the special case
$\lambda {=} \frac{2m^2_{({\rm A})}}{\phi^2_*}$, for which the potential
\begin{equation}
V(\phi^2) \equiv  m^2_{({\rm A})} (\phi^2 {-} \phi^2_*)  + \frac12 \lambda
\left(\phi^2-\phi^2_* \right)^2 = \frac{m^2_{({\rm A})}}{\phi^2_*} \phi^2 (\phi^2 {-} \phi^2_*)
\label{V}
\end{equation}
has a local maximum $V_{({\rm max})} {=}V(0){=}0$ at $\phi{=}0$ and two symmetric minima
$V_{({\min})}{=} {-} \frac14 m^2_{({\rm A})} \phi^2_* $ at $\phi {=}\pm \frac{\phi_*}{\sqrt2}$. As usual, the local maximum is instable.

\subsection{Master equations for the axion electrodynamics}

The set of master equations of axion electrodynamics can be divided into three sub-groups:
first, the evolutionary equations for the Maxwell tensor $F_{ik}$; second, the equation for the pseudo-scalar field $\phi$;
third, equations for the medium dynamics, describing the evolution of the macroscopic velocity four-vector $U^i$ and the energy balance.

The variation of the action functional (\ref{actmin}) with respect
to the four-vector potential $A_i$ gives the electrodynamic
equations
\begin{equation}
\nabla_{k} H^{ik} = -I^i \,,  \label{R2}
\end{equation}
where
\begin{equation}
H^{ik} \equiv C^{ikmn}F_{mn} +  \phi F^{*ik}
\label{R21}
\end{equation}
is the excitation tensor, and
\begin{equation}
I^i \equiv \frac{\delta L_{({\rm m})}}{\delta A_i}
\label{R211}
\end{equation}
is the electric current four-vector.
From Eqs. (\ref{Emaxstar}), (\ref{Cikmn}) and the definition (\ref{R21}), the Eq. (\ref{R2})
can be transformed into
\begin{equation}
\nabla_{k} \left[F^{ik} + (n^2{-}1)\left(F^{im}U^k - F^{km}U^i
\right)U_m \right] = - \mu \left(F^{*ik} \nabla_k \phi + I^i \right) \,.
\label{Emaxmin1}
\end{equation}
Equations for the axion field can be obtained from the action
(\ref{actmin}) by the variation with respect to the pseudoscalar
field $\phi$, yielding
\begin{equation}
\left[\nabla^k  \nabla_k {+} m^2_{({\rm A})} {+} \lambda (\phi^2
{-} \phi^2_*)\right] \phi = - \frac{1}{4 \Psi^2_0} F^{*mn}F_{mn}
\,. \label{induc3}
\end{equation}
Equations of axion electrodynamics have to be supplemented by equations for the gravitational field and by evolutionary equations for the velocity four-vector $U^i$.

\subsection{Master equations for the gravitational field}

Variation of the action functional (\ref{actmin}) with respect to metric gives the equations for the gravitational field, which have the standard form:
\begin{equation}
R_{ik}-\frac12 g_{ik} R = \kappa T^{({\rm eff})}_{ik} \,. \label{GR}
\end{equation}
Here the effective stress-energy tensor $T^{({\rm eff})}_{ik}$ contains three distinguished parts
\begin{equation}
T^{({\rm eff})}_{ik}= T^{({\rm matter})}_{ik} + T^{({\rm A})}_{ik} + T^{({\rm EM})}_{ik} \,. \label{u2}
\end{equation}
The term $T^{({\rm matter})}_{ik}$ is the stress-energy tensor of the matter defined as
\begin{equation}
T^{({\rm matter})}_{ik}= - \frac{2}{\sqrt{-g}}\frac{\delta}{\delta g^{ik}} \left[\sqrt{-g} L_{({\rm matter})}\right] \,. \label{u3}
\end{equation}
The stress-energy tensor of the pseudoscalar field, $T^{({\rm A})}_{ik}$, is of the form
\begin{equation}
T^{({\rm A})}_{ik} \equiv \Psi^2_0 \left\{\nabla_i \phi \nabla_k
\phi - \frac{1}{2} g_{ik} \left[ \nabla^m \phi
\nabla_m \phi - m^2_{({\rm A})} (\phi^2 {-} \phi^2_*)  - \frac12 \lambda
\left(\phi^2-\phi^2_* \right)^2\right] \right\}
\,. \label{TAX}
\end{equation}
The last term, $T^{({\rm EM})}_{ik}$, is the effective stress-energy tensor of the electromagnetic field in the medium defined as
\begin{equation}
T^{({\rm EM})}_{ik} = - F_{ab} F_{pq} \frac{2}{\sqrt{-g}}\frac{\delta}{\delta g^{ik}} \left[\sqrt{-g} C^{abpq} \right] \,. \label{u7}
\end{equation}
Let us remind that the term $\frac{1}{4} \phi \sqrt{{-}g} F^{*mn}F_{mn} {=} \frac{1}{8} \phi E^{ikmn} F_{ik}F_{mn}$
does not depend on the metric and thus does not contribute the effective stress-energy tensor.

In order to represent the matter stress-energy tensor defined formally by the relationship (\ref{u3}), we use the standard procedure based on the Landau-Lifshitz definition of the macroscopic velocity four-vector $U^i$. It can be introduced as the time-like unit  eigen-vector of this tensor and has to satisfy the following equalities:
\begin{equation}
T^{({\rm matter})}_{ik} U^k = W U_i \,, \quad U^i U_i=1 \,. \label{u6}
\end{equation}
Using this unit four-vector we can decompose  the tensor $T^{({\rm matter})}_{ik}$ as
\begin{equation}
T^{({\rm matter})}_{ik} = W U_i U_k + {\cal P}_{ik} \,, \label{u4}
\end{equation}
where $W$ is the matter energy density and ${\cal P}_{ik}$ is the pressure tensor  given by
\begin{equation}
W \equiv U^i T^{({\rm matter})}_{ik} U^k \,,  \quad {\cal P}_{ik} \equiv  \Delta^p_i T^{({\rm matter})}_{pq} \Delta^q_k \,. \label{u5}
\end{equation}
As usual, $\Delta^p_i \equiv \delta^p_i {-} U^p U_i$ is the projection tensor.
The stress-energy tensor (\ref{u7}) with $C^{abpq}$ given by (\ref{Cikmn}) can be written in the form
\begin{equation}
T^{({\rm EM})}_{ik} = \frac14 g_{ik} C^{mlpq} F_{ml}F_{pq} - \frac12 \left(C_{i}^{\ mpq} F_{km} + C_{k}^{\ mpq} F_{im} \right)F_{pq}  \,, \label{u8}
\end{equation}
(see, e.g., \cite{CQG2007,GC2007} for details).
This tensor is symmetric, traceless and coincides with the
symmetrized Minkowski energy - momentum tensor of the
electromagnetic field in the moving medium. Clearly, this tensor
coincides with the standard vacuum stress-energy tensor if we put
$n^2 {=}1$ and $\mu {=}1$ into (\ref{Cikmn}).

\subsection{Evolutionary equations for the macroscopic velocity}

The Bianchi identities require the total stress-energy tensor $T^{({\rm eff})}_{ik} $ to be  divergence-free, i.e.,
\begin{equation}
\nabla^k  \left( T^{({\rm matter})}_{ik} + T^{({\rm A})}_{ik} + T^{({\rm EM})}_{ik} \right) = 0  \,. \label{u11}
\end{equation}
For the axion part of this tensor we obtain that
\begin{equation}
\nabla^k T^{({\rm A})}_{ik} = \Psi^2_0 \nabla_i \phi \left[\nabla^m \nabla_m + V^{\prime}(\phi^2) \right]\phi =
 - \frac14 \nabla_i \phi F^*_{mn}F^{mn}\,, \label{u111}
\end{equation}
thus, taking into account Eqs. (\ref{R2}) and (\ref{R21}), and the identity
\begin{equation}
F^{im}F^*_{km} =
 \frac14 \delta^i_k F^*_{mn}F^{mn}\,, \label{u121}
\end{equation}
(see, e.g., Appendix A in \cite{BMZ}),
we see that for a currentless medium the divergence of the axion stress-energy tensor is of the form
\begin{equation}
\nabla^k T^{({\rm A})}_{ik} = F_{im} \nabla_k (C^{kmpq}F_{pq}) \,. \label{u112}
\end{equation}
We use the standard decomposition
\begin{equation}
{\cal P}_{ik} = - P \Delta_{ik} + \Pi_{ik} \,, \label{u113}
\end{equation}
where $P$ is the Pascal pressure scalar, and  $\Pi_{ik}$ is the non-equilibrium pressure tensor, and then project
the equation (\ref{u11}) onto the direction $U^i$ and the surface orthogonal to it. The corresponding scalar equation
\begin{equation}
DW+(W+P)\Theta = \Pi^{ik} \nabla_k U_i + \Gamma
\label{u114}
\end{equation}
describes the energy balance in the system. Here $D{=}U^i\nabla_i$ is the convective derivative; $\Theta {=} \nabla_i U^i$
is the expansion scalar, and $\Gamma \equiv - U^i {\cal T}_i$. The four-vector ${\cal T}_i$ introduces the so-called ponderomotive force
\begin{equation}
{\cal T}_i = \frac12 \left[F_{im} \nabla_k M^{km} {-}  M_{im} \nabla_k F^{km} \right] {+}
\frac14 F^{pq}\left[\nabla_i M_{pq}+\nabla_p M_{qi} {+} \nabla_q M_{ip} \right] \,,
\label{u115}
\end{equation}
where $M_{ik}$ is the polarization-magnetization tensor defined as follows:
\begin{equation}
M^{ik} =  C^{ikmn} F_{mn} - F^{ik}  \,.
\label{u125}
\end{equation}
The scalar $\Gamma$ describes the contribution of electromagnetic field into the energy balance.
Convolution of (\ref{u11}) with projector $\Delta^{li}$ gives
\begin{equation}
(W+P)DU^l = \Delta^{lk}\nabla_k P - \Delta^{li} \nabla^k \Pi_{ik} + \Gamma^l \,,
\label{u119}
\end{equation}
where $\Gamma^l \equiv - \Delta^{li} {\cal T}_i$.
Eq. (\ref{u119}) is the evolutionary equation for the macroscopic velocity; it is of the first order in convective derivative $DU^l$,
and of the second order in the spatial derivatives $\Delta^{k}_i\nabla_k U_m $ in case when the non-equilibrium pressure tensor $\Pi_{pq}$
is of Navier-Stokes form. Thermodynamic contributions to this equations can be taken into account, e.g., using the scheme discussed in \cite{tds}.
This evolutionary equation depends on the Maxwell tensor via the term $\Gamma^l$, but does not depend explicitly on the axion field and its derivatives.

\section{Electrodynamic system coupled to axion field \\ in the pp-wave gravitational background}

\subsection{PP-wave gravitational background}

We consider the background space-time with pp-wave symmetry in the
so-called TT-gauge (transverse - traceless). The line element,
which we use below
\begin{equation}
\mbox{d}s^{2} = 2\mbox{d}u\mbox{d}v - L^{2} \left[e^{2
\beta}(\mbox{d}x^2)^2 + e^{-2\beta} (\mbox{d}x^3)^2 \right] \,,
\label{metric}
\end{equation}
describes the gravitational pp-wave of the first polarization (the so-called plus polarization)
(see, e.g., \cite{MTW}). Here $u {=} \frac{ct {-} x^1}{\sqrt{2}}$ is the retarded time,
$v {=} \frac{ct {+} x^1}{\sqrt{2}}$ is the advanced time, and $L(u)$, $\beta(u)$ are the functions of the retarded time only.
The pp-wave metric (\ref{metric})
admits the following set of Killing vector fields:
$$
\xi^{i}_{(v)}= \delta ^{i}_{v}\,,  \quad
\xi^{i}_{(2)}= \delta ^{i}_{2}\,,  \quad
\xi^{i}_{(3)}= \delta ^{i}_{3}\,,
$$
\begin{equation}
\xi^{i}_{(4)}= x^{2}  \delta^{i}_{v}  +  \delta^{i}_{2}
\int{L^{-2}(u) e^{-2\beta(u)}du}\,, \quad
\xi^{i}_{(5)} = x^{3}  \delta^{i}_{v}
+\delta^{i}_{3}  \int{L^{-2}(u) e^{2\beta(u)} du} \,,
                                                 \label{killings}
\end{equation}
i.e., the Lie derivative of the metric along these Killing vectors
$\xi_{(a)}$ is equal to zero $\pounds_{\xi_{(a)}} g_{ik} {=} 0$.
Therefore, the metric (\ref{metric})
possesses $G_5$ as the symmetry group
\cite{ExactSolutions}.  The first three Killing vectors, $\xi^{i}_{(v)}$, $\xi^{i}_{(2)}$, and
$\xi^{i}_{(3)}$, form a $G_3$ Abelian subgroup of $G_5$. The vector $\xi^{i}_{(v)}$ is
isotropic, covariantly constant and orthogonal to the other four ones,
i.e.,
\begin{equation}
\nabla_{k} \  \xi^{i}_{(v)}=0\,, \quad
g_{ik} \  \xi^{i}_{(v)}  \xi^{k}_{(j)} =0 \,.
\label{orthogonality}
\end{equation}
Let ${\bf \Psi}$ be an arbitrary macroscopic function of the state
of the system (material tensor, Maxwell tensor, induction tensor,
pseudoscalar (axion) field, etc.). When the quantity  ${\bf \Psi}$
as the solution of master equations of the model satisfies the
conditions ${\pounds}_{\xi_{(2)}} {\bf \Psi} =0$ and
${\pounds}_{\xi_{(3)}} {\bf \Psi} =0$, we can state that it
inherits the {\it plane} symmetry supported by the gravitational
wave field. One obtains in this case that ${\bf \Psi}$ does not
depend on variables $x^2$ and $x^3$, being the function of $u$ and
$v$ only. Let us imagine that the solution of master equations
satisfies an additional condition ${\pounds}_{\xi_{(v)}} {\bf
\Psi} =0$, i.e., we deal with three relationships
${\pounds}_{\xi_{(b)}} {\bf \Psi} =0$ for all three Killing
vectors $\xi^i_{(b)}$ ($b{=}v,2,3$) belonging to the Abelian
subgroup $G_3$ of the total $G_5$ group. Then we can indicate the
corresponding field or state function as inheriting the {\it
plane-wave} symmetry of the GW background (see, e.g.,
\cite{BL,BN10}), and consider ${\bf \Psi}$ as a function of the
retarded time $u$ only.

The initial data for the metric functions on the null
hyper-surface $u{=}0$ can be formulated as follows
\begin{equation}
L(0)=1 \,, \quad L^{\prime}(0) = 0 \,, \quad \beta(0)=0 \,, \quad
\beta^{\prime}(0) = 0 \,. \label{initials}
\end{equation}
Similarly, we use below the terms $\phi(0)$ and $F_{ik}(0)$ as the initial data for the axion field and for
the electromagnetic field, respectively, fixed on the  null
hyper-surface $u{=}0$.

In this paper we consider the so-called {\it test} electromagnetic and pseudoscalar
fields, i.e., the gravitational pp-wave field is assumed to be unperturbed
by these fields, or in other words, the curvature introduced by
these test fields are negligible in comparison with the curvature
produced by incoming (background) gravitational waves. Corresponding estimations,
which restrict our prognosis for the case of weak pp-wave from a periodic astrophysical source, are quoted below in Section 6.2.3.

\subsection{Initial state}

We assume that at $u<0$, i.e., before the gravitational wave (GW,
for short) appearance the magnetic field was characterized by
three constant components $B^{(1)}$, $B^{(2)}$ and $B^{(3)}$
related to the coordinate system attributed to the GW field
(\ref{metric}). To be more precise, we consider the axis $Ox^1$ to
be the direction of the GW propagation, thus $B^{(1)}$ can be
classified as a longitudinal component. We indicate two
eigen-directions in the field of GW as $Ox^2$ and $Ox^3$,
respectively. In this sense it is convenient to use the following
definitions
\begin{equation}
B^{(2)} = B_{\bot} \cos{\Theta} \,, \quad B^{(3)} = B_{\bot}
\sin{\Theta} \,,
 \label{sol7}
\end{equation}
for two transversal components of the magnetic field, where
$\Theta$ plays the role of azimuthal angle in the plane of the GW
front, $x^2Ox^3$. Initial values of $F_{ik}$ are linked with
magnetic field components by the relations
$$
F_{23} = - B^{(1)} \,, \quad  F_{13} = B^{(2)}  \,, \quad F_{12} =
- B^{(3)} \,,
$$
\begin{equation}
F^{10} = E^{(1)} = 0 \,, \quad F^{20} = E^{(2)} = 0 \,, \quad
F^{30} = E^{(3)}=0 \,, \label{mf1}
\end{equation}
and initial electric field components $E^{(1)}$, $E^{(2)}$,
$E^{(3)}$ are vanishing.

Since the pseudoscalar $I^{*} \equiv \frac14 F^{*}_{mn}F^{mn}$ is
equal to zero for this initial electromagnetic field
configuration, the source-term in the right-hand-side of
(\ref{induc3}) vanishes and there is no coupling between axion and
electromagnetic fields at $u<0$. We assume that the pseudoscalar
field possesses the plane symmetry and thus it does not depend on
$x^2$ and $x^3$. Then we obtain the following evolutionary
equation for the quantity $\phi(u,v)$:
\begin{equation}
\partial_{u}\partial_{v} \ \phi = {\cal F} \,, \quad {\cal F} = - \frac12 \phi \left[ m^2_{({\rm A})} {+} \lambda
\left(\phi^2 {-} \phi^2_* \right)\right]   \,. \label{0set1}
\end{equation}
If we add to this equation the data on the characteristic lines
\begin{equation}
\phi(0,v) = \mu(v) \,, \quad \phi(u,0) = \nu(u) \,, \quad  \mu(0)
= \nu(0) \,, \label{oset1}
\end{equation}
we obtain the particular case of the classical Goursat problem
(see, e.g., \cite{Goursat}), the solution of which is known to
exist and to be unique. We assume here that $\mu(v) {=} \nu(u<0)
{=} \phi(0)$, and $\phi(0)$ satisfies the algebraic equation of
the third order
\begin{equation}
\phi(0) \ \{m^2_{({\rm A})} + \lambda [\phi^2(0) - \phi^2_*]\} = 0
\,. \label{mf111}
\end{equation}
Then one obtains that at $u<0$ the solution of the equation
(\ref{0set1}) is the constant solution $\phi(u\leq 0,v,x^2,x^3)
{=}\phi(0)$. Clearly, one of the solutions of (\ref{mf111}) is
trivial, $\phi(0){=}0$; two other solutions are $\phi(0){=} \pm
\sqrt{\phi^2_* {-} \frac{m^2_{({\rm A})}}{\lambda}}$. When
$\phi^2_* \leq \frac{m^2_{({\rm A})}}{\lambda}$, the trivial
solution $\phi(0) {=}0$ is unique. Finally, we assume that, when
GW is absent, the medium is homogeneous and is in the state of
rest. This means that the medium energy-density is constant, $W(u
\leq 0)=const$, and the velocity four-vector has the form $U^i(u
\leq 0) {=} \delta^i_0$; the equations (\ref{u114})-(\ref{u119})), clearly, admit such solutions at $u<0$.

In other words, we assume that before the GW appearance ($u<0$)
the state of the electrodynamic system coupled to the pseudoscalar
field was static and homogeneous, i.e., the basic quantities
$\phi$ and $F_{mn}$ did not depend on time and spatial
coordinates. Of course, this ansatz assumes that the model system
has no space-like boundaries.

\subsection{Reduced master equations}

Let us consider the coupled system of equations of the axion
electrodynamics in the pp-wave background (i.e., at $u>0$). This
system is reduced from (\ref{induc3}), (\ref{Emaxstar}) and
(\ref{Emaxmin1}), and it contains three sub-systems. The first
sub-system gives us the equation for the pseudoscalar field
\begin{equation}
\left[2\left(\partial_{u}{+}\frac{L^{\prime}}{L}
\right)\partial_{v} {-}
\frac{1}{L^2}\left(e^{-2\beta}\partial^2_{2} {+}
e^{2\beta}\partial^2_{3} \right) {+} m^2_{({\rm A})} {+} \lambda
\left(\phi^2 {-} \phi^2_* \right)\right] \phi = {-} \frac{1}{L^2
\Psi^2_{0}} F_{u(v} F_{23)} \,. \label{set01}
\end{equation}
Here and below the prime indicates the ordinary derivative with
respect to retarded time $L^{\prime} \equiv \frac{dL}{du}$; the
symbol $\partial_k$ stands for the partial derivative; the symbol
$(ijk)$ denotes the cyclic transposition of three mentioned
indices. The second sub-system of equations
\begin{equation}
\partial_{(2}F_{uv)}=0 \,, \quad  \partial_{(3}F_{uv)}=0 \,, \quad \partial_{(2}F_{u3)}=0 \,, \quad \partial_{(2}F_{v3)}=0 \,,  \label{set02}
\end{equation}
comes from (\ref{Emaxstar}). The third sub-system is of the form
$$
\partial_{v}[L^2F_{uv}] {+} e^{{-}2\beta} \partial_{2} \left[\frac{(n^2{+}1)}{2n^2}F_{v2}{+}
\frac{(n^2{-}1)}{2n^2}F_{u2}\right]{+} e^{2\beta}
\partial_{3} \left[\frac{(n^2{+}1)}{2n^2}F_{v3}{+} \frac{(n^2{-}1)}{2n^2}F_{u3}\right]{=}
\frac{\mu}{n^2} F_{(23}
\partial_{v)}\phi,
$$
$$
\partial_{u} \left[L^2 F_{uv}\right] {-} e^{{-}2\beta} \partial_{2}\left[\frac{(n^2{+}1)}{2n^2}F_{u2}{+}
\frac{(n^2{-}1)}{2n^2}F_{v2}\right]  {-} e^{2\beta}
\partial_{3}\left[\frac{(n^2{+}1)}{2n^2}F_{u3}{+}
\frac{(n^2{-}1)}{2n^2}F_{v3}\right] {=} \frac{\mu}{n^2}F_{(23}
\partial_{u)}\phi,
$$
$$
\partial_{u} \left\{ e^{{-}2\beta}\left[\frac{(n^2{+}1)}{2}F_{v2}{+}
\frac{(n^2{-}1)}{2}F_{u2}\right]\right\} {+} e^{{-}2\beta}
\partial_{v}\left[\frac{(n^2{+}1)}{2}F_{u2}{+}
\frac{(n^2{-}1)}{2}F_{v2}\right] {+}\frac{1}{L^2}
\partial_{3} F_{23} {=}\mu F_{(vu}
\partial_{3)}\phi,
$$
\begin{equation}
\partial_{u} \left\{ e^{2\beta}\left[\frac{(n^2{+}1)}{2}F_{v3}{+}
\frac{(n^2{-}1)}{2}F_{u3}\right]\right\} {+} e^{2\beta}
\partial_{v}\left[\frac{(n^2{+}1)}{2}F_{u3}{+}
\frac{(n^2{-}1)}{2}F_{v3}\right] {-}\frac{1}{L^2}
\partial_{2} F_{23} {=} \mu F_{(uv}
\partial_{2)}\phi, \label{set03}
\end{equation}
and is obtained from (\ref{Emaxmin1}) with $i{=}u,v,x^2,x^3$,
respectively. The solutions to this system of equations differ
essentially for the cases $n^2 \equiv 1$ and $n^2 \neq 1$; we
start to analyze exact solutions related to the first case.

\section{Evolution of electromagnetic and axion fields in the gravitational-wave background. The case: $n^2 \equiv 1$}

\subsection{Exact solution to the master equations in the general case: \\ $B^{(1)} \neq 0$, $B^{(2)} \neq 0$ and $B^{(3)} \neq 0$}

Let us consider the coupled system of equations
(\ref{set01})-(\ref{set03}) in the case $n^2{=}1$, $\mu{=}1$. We
omit technical details of the system integration. The Reader can
check directly that the following functions give an exact solution
to this system of equations:
\begin{equation}
\phi(u,v,x^2,x^3) =  \phi(0) - 2 \arctan{
\left[\frac{\sin{2\Theta} \ \sinh{\beta(u)}}{\cosh{\beta(u)} +
\cos{2\Theta} \sinh{\beta(u)}} \right]} \equiv \Phi(u,\Theta) \,,
\label{set5}
\end{equation}
$$
F_{uv}(u,v,x^2,x^3) = - \frac{B^{(1)}}{L^2(u)} [\Phi(u,\Theta) {-}
\phi(0)] \equiv E_{||}(u)\,, $$
\begin{equation}
F_{23}(u,v,x^2,x^3) = - B^{(1)} \equiv B_{||}(0) \,, \label{set6}
\end{equation}
\begin{equation}
F_{v2}(u,v,x^2,x^3) = - \frac{B^{(3)}}{\sqrt2} \ a(u) \,, \quad
F_{v3}(u,v,x^2,x^3) = \frac{B^{(2)}}{\sqrt2} \ a(u)\,,
\label{set10}
\end{equation}
\begin{equation}
F_{u2}(u,v,x^2,x^3) = -\frac{B^{(3)}}{\sqrt2} \left[v a^{\prime}(u) - 1 - b(u) \right]\,, \label{set7}
\end{equation}
\begin{equation}
F_{u3}(u,v,x^2,x^3) = \frac{B^{(2)}}{\sqrt2} \left[v a^{\prime}(u)
- 1 + b(u) \right]\,. \label{set8}
\end{equation}
Here we used the following auxiliary functions:
\begin{equation}
a(u) \equiv \frac{1}{\sqrt{\cosh{2 \beta(u)} + \cos{2\Theta}
\sinh{2\beta(u)}}} \,, \quad a(0) = 1 \,,
 \label{set12}
\end{equation}
\begin{equation}
b(u) \equiv \frac{2 \Psi^2_0 L^2(u)}{a(u) B^2_{\bot}\sin{2\Theta}}
\left[{\cal H}(\Phi) + \frac{(B^{(1)})^2}{L^4 \Psi^2_0} \right]\left[\Phi(u,\Theta) {-} \phi(0) \right]
\,, \quad b(0) = 0 \,,
 \label{set13}
\end{equation}
\begin{equation}
{\cal H}(\Phi) \equiv  m^2_{({\rm A})} + \lambda \left[\Phi^2 +
\Phi \phi(0) + \phi^2(0) - \phi^2_* \right] \,.
 \label{set14}
\end{equation}
Clearly, the solutions for the components of the Maxwell tensor
correspond to the homogeneous initial data obtained from the conditions
(\ref{mf1})
$$
F_{uv}(u{=}0,v,x^2,x^3) = 0 \,, \quad F_{23}(u{=}0,v,x^2,x^3) = -
B^{(1)} \,,
$$
$$
F_{u2}(u{=}0,v,x^2,x^3) = \frac{B^{(3)}}{\sqrt2} \,, \quad
F_{u3}(u{=}0,v,x^2,x^3) = - \frac{B^{(2)}}{\sqrt2} \,,
$$
\begin{equation}
 F_{v2}(u{=}0,v,x^2,x^3) =
-\frac{B^{(2)}}{\sqrt2}\,, \quad F_{v3}(u{=}0,v,x^2,x^3) =
\frac{B^{(3)}}{\sqrt2}\,. \label{set4}
\end{equation}
It is worth mentioning three details of these exact solutions for
the components of the Maxwell tensor at $u>0$. First, $F_{ik}$ do
not depend on $x^2$ and $x^3$, i.e., inherit the {\it plane}
symmetry of the GW field. Second, $F_{u2}$ and $F_{u3}$ are linear
in the advanced time $v$, but other components of the  Maxwell
tensor depend on retarded time $u$ only. Third, the
pseudo-invariant $I^* {=}\frac14 F^*_{mn}F^{mn}$ does not depend
on $v$, i.e., it inherits full {\it plane-wave} symmetry of the GW
field.

Concerning the solution for $\phi$ (see (\ref{set5})), it depends
on the retarded time $u$ only, and satisfies the conditions
\begin{equation}
\Phi(0, \Theta) = \phi(0) \,, \quad \Phi^{\prime}(0,\Theta)=0 \,.
\label{set41}
\end{equation}
Clearly, this exact solution can be interpreted as inheriting the
plane-wave symmetry of the GW field. Let us briefly discuss the
problem of uniqueness of the solution $\Phi(u,\Theta)$ starting
from an obvious assumption that $\phi$ depends on $u$ and $v$ only
(the axion field inherits the plane symmetry, but the plane-wave
symmetry is not obligatory). Then the evolutionary equation for
$\phi(u,v)$ can be rewritten in the form
\begin{equation}
\partial_{u} \partial_{v} \phi =
\tilde{{\cal F}}(u,v,\phi, \partial_{u}\phi, \partial_{v} \phi) \,, \label{set991}
\end{equation}
with function $\tilde{{\cal F}}$ in the right-hand side, which
contains now the partial derivatives of the first order,
$\partial_{u}\phi$ and $\partial_{v} \phi$, since the
pseudo-invariant $I^*$ is not vanishing. When we add to this
equation the conditions on the characteristics
\begin{equation}
\phi(0,v) = \mu(v) \,, \quad \phi(u\geq 0,0) = \tilde{\nu}(u) \,,
\quad \mu(0) = \tilde{\nu}(0) = \phi(0) \,, \label{set199}
\end{equation}
we obtain again the Goursat problem for $u>0$ associated with the
one considered above for $u<0$. It is well-known that the solution
of the Goursat problem exists and is unique, in particular, when
the function $\tilde{{\cal F}}$ satisfies the Lipschitz conditions
with respect to $\phi$, $\partial_{u}\phi$ and $\partial_{v}
\phi$. In our case the function $\tilde{{\cal F}}$ satisfies these
Lipschitz conditions, when $\sin{2\Theta}\neq 0$, i.e., when $B^{(2)} \cdot
B^{(3)} \neq 0$. If we put again $\mu(v){=}\phi(0)$ and
$\tilde{\nu}(u){=}\Phi(u,\Theta)$, we can state that the solution
for $\phi$, which we presented in (\ref{set5}), is unique.

In other words, we proved that the vacuum ($n^2{=}1$) model of
coupling of pseudoscalar (axion) field with initially constant
magnetic field in the pp-wave background admits exact solution,
for which the axion field happens to inherit the pp-wave symmetry of the GW
background.

\subsection{Special solutions  with $B^{(2)} \cdot B^{(3)} {=} 0$}

Since the solution presented above is not defined at
$\sin{2\Theta}{=}0$ (see (\ref{set13})), we focus a special
attention on the case when $B^{(2)} \cdot  B^{(3)} {=} 0$. This
special case can be divided into two sub-cases.

\subsubsection{Pure longitudinal magnetic field ($B^{(1)} \neq 0$ and  $B^{(2)} {=} B^{(3)} {=} 0$)}

It is very easy to check directly that the following functions:
\begin{equation}
\phi =  \phi(0) \,,  \quad F_{23} = - B^{(1)}  \,, \label{set66}
\end{equation}
\begin{equation}
F_{uv} = 0 \,, \quad F_{v2} = 0 \,, \quad F_{v3} = 0 \,, \quad
F_{u2} = 0 \,, \quad F_{u3} = 0 \,. \label{set88}
\end{equation}
satisfy the equations (\ref{set01})-(\ref{set03}). The GW does not
initiate any changes in such configuration of the electrodynamic
system.

\subsubsection{Transversal magnetic field ($B^{(3)} \neq 0$ and $B^{(1)} = B^{(2)} = 0$)}

For such initial configuration the solution to
(\ref{set01})-(\ref{set03}) is
\begin{equation}
\phi =  \phi(0) \,, \quad  F_{v2} = - \frac{B^{(3)}}{\sqrt2} \
e^{\beta(u)}\,, \quad F_{u2} = \frac{B^{(3)}}{\sqrt2} \left[1- v
\beta^{\prime}(u) e^{\beta(u)} \right] \,, \label{set77}
\end{equation}
\begin{equation}
F_{uv} = 0 \,, \quad F_{23} = 0  \,, \quad F_{v3} = 0 \,, \quad
F_{u3} = 0 \,. \label{set89}
\end{equation}
Thus, the distortion of the magnetic field is produced by the
gravitational wave only, and there is no effects induced by
axion-photon interaction.

\subsection{GW- induced distortion of the initial magnetic field}

\subsubsection{Axionic contribution to the distortion of the initial magnetic field
in the GW-background}

Now we present the formulas describing the changes in the state of
electromagnetic field induced both by gravitational wave and axion
field. We deal with the so-called physical components of magnetic
and electric field defined as
$$
{\cal B}^1 \equiv B^1 \,, \quad  {\cal B}^2 \equiv
\sqrt{-g_{22}(B^2)^2} \,, \quad {\cal B}^3 \equiv
\sqrt{-g_{33}(B^3)^2}  \,,
$$
\begin{equation}
{\cal E}^1 \equiv E^1 \,, \quad {\cal E}^2 \equiv
\sqrt{-g_{22}(E^2)^2} \,, \quad {\cal E}^3 \equiv
\sqrt{-g_{33}(E^3)^2} \,.
 \label{defin}
\end{equation}
The results are the following. First, the longitudinal magnetic
field $B^{1}$ coincides with its physical component ${\cal B}^1$
and is not distorted. Second, the longitudinal electric field
$E^{1}$ coincides with its physical component and contains the
axion part of distortion only
\begin{equation}
 {\cal E}^1(u) = E^1(u)  =   \frac{2B^{(1)}}{L^2} \arctan{ \left[\frac{\sin{2\Theta}
\ \sinh{\beta}}{\cosh{\beta} + \cos{2\Theta} \sinh{\beta}}
\right]}\,.
 \label{phys7}
\end{equation}
Third, the transversal quantities can be divided into three and two parts, respectively:
\begin{equation}
{\cal B}^2 = L e^{\beta} B^{(2)} \left[1 + X(u,v) + Z(u)\right]
 \,,  \quad
 {\cal B}^3 = L e^{-\beta} B^{(3)} \left[ 1 + X(u,v)  - Z(u)\right] \,,
 \label{phys2}
\end{equation}
\begin{equation}
 {\cal E}^2 = \frac{B^{(3)}}{L} e^{\beta}  \left[ - Y(u,v)  - Z(u)\right]
 \,, \quad
 {\cal E}^3 = \frac{B^{(2)}}{L} e^{- \beta} \left[ Y(u,v)  - Z(u)\right]
 \,.
 \label{phys4}
\end{equation}
Here two dimensionless distortion functions, defined as
\begin{equation}
 X(u,v) = \frac{1}{2} \left[ a(u) -1 - va^{\prime}(u) \right] \,,
 \quad Y(u,v) = \frac{1}{2} \left[ a(u) -1 + va^{\prime}(u) \right]
 \,,
 \label{phys5}
\end{equation}
describe pure gravitational wave influence on the magnetic field,
since there is no information about axion field in these terms.
The third function
\begin{equation}
 Z(u) =  \frac{2 \Psi^2_0 L^2}{a(u) B^2_{\bot}\sin{2\Theta}}
\left[{\cal H}(\Phi) {+} \frac{(B^{(1)})^2}{L^4 \Psi^2_0} \right]
\arctan{ \left[\frac{\sin{2\Theta} \ \sinh{\beta}}{\cosh{\beta}
{+} \cos{2\Theta} \sinh{\beta}} \right]} \,,
 \label{phys6}
\end{equation}
introduces the distortion caused by the interaction with axion
field. All the supplementary expressions vanish at $u<0$, as it
should be. Since in this model we deal with (axionic) vacuum, the
velocity four-vector, $U^i$, does not appear in the formulas for
the electromagnetic response. Thus, we presented a {\it new exact
solution} to the equations of axion electrodynamics in the
gravitational pp-wave background, which generalizes the solution
obtained in \cite{BL}.

\subsubsection{Anomalous character of the electromagnetic response}

The function $Z(u)$ (\ref{phys6}) involves into discussion a principally new term describing an anomaly
in the electromagnetic response on the gravitational wave action, which is formed in the axionic dark matter environment.
Indeed, when $\Psi^2_0 m^2_{({\rm A})} \neq 0$, $\beta(u) \neq 0$ and $B_{\bot} \neq 0$, the function $Z(u)$ contains
$B^2_{\bot} \neq 0$ in the denominator.
This means that at presence of the gravitational wave the electromagnetic response grows anomalously, when $B^2_{\bot} \to 0$.
Nevertheless, when $B^2_{\bot} \equiv 0$
identically, the effect vanishes. Clearly, we deal with critical behavior of the response,
since $\lim_{B_{\bot}\to 0}\{F_{ik}(B_{\bot}) \} {=} \infty \neq \{F_{ik}(B_{\bot}{=}0) \}$, when $\beta \neq 0$.
Moreover, even  if (hypothetically) $m_{({\rm A})}{=}0$, i.e., axions are assumed to be massless, and the longitudinal
magnetic field is non-vanishing, i.e., $B^{(1)}\neq 0$, the term $\left(\frac{B^{(1)}}{B_{\bot}}\right)^2$ displays the same
critical behavior at $B_{\bot} \to 0$.

\section{Evolution of electromagnetic and axion fields in the gravitational-wave background. The case: $n^2 \neq 1$}

\subsection{Exact solutions}

Let us consider now the medium with $n^2 \neq 1$. The Reader can
check directly that the following functions:
\begin{equation}
F_{uv}(u) = - \frac{B^{(1)}}{\varepsilon
L^2}\left[\phi(u){-}\phi(0) \right] \,, \quad F_{23}(u) = -
B^{(1)}\,,  \label{h3}
\end{equation}
\begin{equation}
F_{v2}(u) = -\frac{B^{(2)}}{\sqrt2}\,, \quad F_{v3}(u) =
\frac{B^{(2)}}{\sqrt2} \,, \label{h2}
\end{equation}
\begin{equation}
F_{u2}(u) = \frac{1}{\sqrt2} e^{2\beta} \left\{B^{(3)} +
\frac{1}{n^2{-}1} \left[(n^2{+}1) B^{(3)} \left(e^{{-}2\beta} {-}1
\right) - 2\mu B^{(2)}\left[\phi(u){-}\phi(0) \right]
\right]\right\} \,, \label{h4}
\end{equation}
\begin{equation}
F_{u3}(u) = {-}\frac{1}{\sqrt2} e^{{-}2\beta} \left\{B^{(2)} +
\frac{1}{n^2{-}1} \left[(n^2{+}1) B^{(2)} \left(e^{2\beta}{-}1
\right) + 2\mu B^{(3)}\left[\phi(u){-}\phi(0) \right]
\right]\right\} \,, \label{h5}
\end{equation}
satisfy the master equation (\ref{set01})-(\ref{set03}), if the
axion field $\phi$ is the solution of algebraic equation of the
third order
$$
{-}\Psi^2_0 L^2 \left[m^2_{({\rm A})} {+} \lambda (\phi^2
{-} \phi^2_*)\right] \phi =
$$
\begin{equation}
= \left[\phi(u){-}\phi(0)
\right]\left\{\frac{{B^{(1)}}^2}{\varepsilon L^2} +
\frac{\mu}{(n^2{-}1)}\left[e^{2\beta} {B^{(2)}}^2 + e^{{-}2\beta}
{B^{(3)}}^2\right] \right\} + \frac{2}{(n^2{-}1)}B^{(2)} B^{(3)}
\sinh{2\beta} \,. \label{induc37}
\end{equation}
In order to illustrate the properties of solutions to this
equation we restrict our-selves by the simplest model with
$\lambda{=}0$ and $\phi(0){=}0$, respectively. Then we obtain
immediately the following expression
\begin{equation}
\phi(u) = {-} \frac{\varepsilon L^2 B^2_{\bot} \sin{2\Theta}
\sinh{2\beta}}{(n^2{-}1)\left(\varepsilon \Psi^2_0 L^4 m^2_{({\rm
A})} {+} {B^{(1)}}^2 \right) {+} n^2 L^2
B^2_{\bot}\left(\cosh{2\beta} {+}\cos{2\Theta} \sinh{2\beta}
\right)} \equiv \Phi(u, \Theta, n)\label{induc38}
\end{equation}
for the pseudoscalar (axion) field. As in the previous case, the
solution of the corresponding Goursat problem is unique, and again
the solution happens to inherit the plane-wave symmetry supported
by the GW background. Let us remark that when $n^2 \neq 1$ all the
components of the Maxwell tensor happen to be functions of
retarded time only, thus, the solutions to electrodynamic
equations inherit the plane-wave symmetry, ${\pounds}_{\xi_{(b)}}
F_{ik} =0$, $b{=}v,2,3$. Moreover, it is clear, that
$F_{ik}\xi^i_{(a)}\xi^k_{(b)} {=} const$, $a \neq b{=}v,2,3$,
thus, only three components $F_{uv}$, $F_{u2}$ and $F_{u3}$ evolve
with retarded time.

\subsection{Anomalous response of the electromagnetic field coupled to the axion field on the gravitational wave action}

When $\varepsilon \to 1$ and $\mu \to 1$ and thus $n^2 \to 1$
the axion field (\ref{induc38}) behaves regularly as
\begin{equation}
\phi(u) \to  - \frac{\sin{2\Theta}
\sinh{2\beta}}{\left(\cosh{2\beta} +\cos{2\Theta} \sinh{2\beta}
\right)} \,.\label{induc381}
\end{equation}
As for the electromagnetic field, we face with a principally
another situation. Indeed, the physical components of the magnetic
and electric fields can be now written as follows. Longitudinal
components of the magnetic and electric fields
\begin{equation}
{\cal B}^1(u) = \frac{B^{(1)}}{L^2} \,, \quad {\cal E}^1(u) = -
\frac{B^{(1)}}{\varepsilon L^2} \phi(u) \,,
 \label{2phys7}
\end{equation}
remain regular, while the transversal components
\begin{equation}
{\cal B}^2 = \frac{1}{L}\left\{B^{(2)}\left[\cosh{\beta} +
\left(\frac{n^2{+}1}{n^2{-}1} \right)\sinh{\beta}  \right] +
\frac{\mu \phi(u)}{(n^2{-}1)} e^{-\beta}B^{(3)}\right\} \,,
 \label{2phys1}
\end{equation}
\begin{equation}
{\cal B}^3 = \frac{1}{L}\left\{B^{(3)}\left[\cosh{\beta} -
\left(\frac{n^2{+}1}{n^2{-}1} \right)\sinh{\beta}  \right] -
\frac{\mu \phi(u)}{(n^2{-}1)} e^{\beta}B^{(2)}\right\} \,,
 \label{2phys2}
\end{equation}
\begin{equation}
 {\cal E}^2 = -\frac{1}{L (n^2{-}1)}\left[2 B^{(3)} \sinh{\beta} + \mu \phi(u) e^{\beta} B^{(2)} \right]
 \,,
 \label{2phys3}
\end{equation}
\begin{equation}
 {\cal E}^3 = -\frac{1}{L (n^2{-}1)}\left[2 B^{(2)} \sinh{\beta} + \mu \phi(u) e^{-\beta} B^{(3)} \right]
 \,,
 \label{2phys4}
\end{equation}
contain irregular parts proportional to the multiplier
$(n^2{-}1)^{-1}$. The first invariant of the electromagnetic field
$I_{1} \equiv \frac14 F_{mn}F^{mn}$, which takes the form
\begin{equation}
I_{1} = \frac{{B^{(1)}}^2}{2\varepsilon L^4}
\left[\varepsilon{-}\phi(u) \right] {+} \frac{B^2_{\bot}}{2L^4}
\left[(\cosh{2\beta}{+} \cos{2\Theta} \sinh{2\beta}) {+}
\frac{4\sinh{\beta}}{(n^2{-}1)}(\sinh{\beta}{+} \cos{2\Theta}
\cosh{\beta}\right]
 \,,
 \label{inv1}
\end{equation}
also contains the term proportional to $(n^2{-}1)^{-1}$.

Studying the behavior of the exact solutions
(\ref{2phys1})-(\ref{2phys4}) we see two distinct situations.
First, if we take, e.g., the term
$\{\frac{\sinh{\beta}}{(n^2{-}1)}\}$ and calculate the limit of
$\beta \to 0$ and then $n^2\to 1$ we obtain that this double limit
is equal to zero, $\lim_{n^2 \to 1} \lim_{\beta \to
0}\{\frac{\sinh{\beta}}{(n^2{-}1)}\} {=}0$. Second, if we take
first the limit of $n^2\to 1$ and then $\beta \to 0$, the double
limit $ \lim_{\beta \to 0} \lim_{n^2 \to 1}
\{\frac{\sinh{\beta}}{(n^2{-}1)}\} {=}\infty$ gives infinity.
Since these two double limits do not coincide, we can speak of a
critical behavior of the electromagnetic field near the singular
point $n^2{=}1$. In the absence of the gravitational wave, i.e., when $\beta
\equiv 0$, such a problem does not arise.

\subsection{Short summary of exact solutions}

Before discussing of the model applications let us summarize features of the obtained exact solutions to the equations
(\ref{set01})-(\ref{set03}), which we presented in Section 4 and Section 5. What is the {\it difference} between these solutions? First of all, the solutions (\ref{set5})-(\ref{set14}) relate to a case, describing an electrically neutral non-conductive medium with unit refraction index, $n{=}1$, i.e., in fact, to a vacuum, in which there are neither atoms (composed of electrically charged particles), nor virtual pairs of particles created by the axion-photon coupling. The solutions (\ref{h3})-(\ref{induc38}) relate to the case, describing a medium with non-unit refraction index, i.e., the presence of residual atoms and/or axionically induced virtual pairs are admissible. Second, the electric and magnetic fields at $n^2 {=}1$ depend on both retarded and advanced times, while at $n^2 \neq 1$ the solutions depend on the retarded time only. Third, only the  solution with $n^2{=}1$ contains the resonance-type part linear in the advanced time. What are {\it similar} details in these solutions? First, the electric field, which was absent initially, appears in both cases under the influence of the GW-field; the initial magnetic field is also distorted in both cases. Second, the axion field in both cases inherits the plane-wave symmetry of the GW field, since in both cases the pseudo-invariant $I^*$ is the function of the retarded time only. Third, in both cases we can find symptoms of critical behavior in the evolution of the electromagnetic filed. Finally, let us emphasize that the anomalous growth of the electric and magnetic field has different features, when we compare the solutions with unit and non-unit refraction indices. When $n^2{=}1$, the anomaly in the GW-induced electromagnetic signal is predetermined by the term (\ref{set13}), which is quadratic in the coupling constant $\Psi_0$ attributed to the axion field and contains the square of initial magnetic field in the denominator; in other words, this anomaly is apparently connected with the axion-photon coupling. When $n^2 \neq 1$, the anomaly appears due to the smallness of a total succeptibility parameter $\chi{=}n^2{-}1$, $\chi{=}\chi_0 {+} \chi_{({\rm axion})}$ located in the denominator. If the residual atoms are absent in the medium (i.e., $\chi_0{=}0$), but $\chi_{({\rm axion})} \neq 0$, again the anomaly appears due to the axion-photon coupling.

\subsection{Analogy with a phase transition of the second kind}

The gravitational pp-wave appearance (at the moment $u{=}0$) can be considered as a specific (space-time) phase
transition of the second kind \cite{Amaldi1,Amaldi2}. It is well-known (see, e.g., \cite{LL5}) that the phase
transition of the second kind can be generally characterized by the change of intrinsic symmetry of the medium. In the context of our model before the
front of the gravitational pp-wave  the space-time has the
symmetry group $G_{10}$ (the so-called symmetric phase), while behind the front the space-time is
described by the group $G_5$ (dissymmetric phase). Clearly, five Killing vectors happen to be lost behind the GW-front. From the physical point of view, one of the typical symptoms
of such phase transitions of the second kind is the creation of new structures in the dissymmetric phase, e.g., spontaneous electric polarization, magnetization and/or deformation in crystals in the vicinity of the corresponding Curie temperature $T_{({\rm C})}$ \cite{LL5}. In the process of transition through the Curie temperature  a number of state functions experience a jump, and their behavior can be characterized by the factors $[T{-}T_{({\rm C})}]^{{-}\gamma}$ with the so-called critical index $\gamma$. When we deal with the impact of the GW-front at $u{=}0$, we see that the internal symmetry of the electrodynamic system is changed accordingly: instead of pure constant magnetic field (in the symmetric phase with $G_{10}$ group) we obtain (as an exact solution) the magnetic field plus electric field (in the dissymmetric phase with $G_5$ group), so that the GW-induced electric field can be indicated as the spontaneous one. Extending this analogy to the case of axion electrodynamics we can compare two terms in the equation (\ref{set01}): the term $m^2_{({\rm A})} \phi$ in the left-hand side, and the electromagnetic source in the right-hand side, which can be reduced to $\frac{1}{L^2 \Psi^2_{0}} (\vec{E} \cdot \vec{B}) $. In the symmetric phase $\vec{E}{=}0$ and static equation for the axion field is satisfied by the solution $\phi{=}0$. In the dissymmetric phase $\vec{E} \neq 0$ and $\phi \neq 0$, thus the modulus of the electric field can be estimated as $|\vec{E}| \propto \frac{m^2_{({\rm A})} \phi L^2
\Psi^2_{0}}{|\vec{B}|}$. The exact solution (\ref{set5})-(\ref{set14}) confirms this reasoning, thus, we can indeed interpret this solution in terms of phase transition of the second kind, and critical behavior of the electric and magnetic field as a natural symptom of this phenomenon.

\section{Applications}

\subsection{Weak gravitational waves and search for amplification of the response signal}

We obtained a new exact solution to the self-consistent set of equations of the axion electrodynamics in
the background of strong gravitational pp-wave with the first polarization. In fact, the experimentalists
are interested in the analysis of effects linear in the weak gravitational wave amplitude $2\beta_0$, appeared due to the typical representation:
\begin{equation}
2\beta(u) = 2\beta_{0}\cos{\left(\frac{\sqrt2}{c} \omega_0 u {+}\psi_0 \right)} =  2\beta_{0}\cos{\left[\omega_0 \left(t-\frac{x^1}{c} \right){+}\psi_0 \right]} \,,
\label{Dphys17}
\end{equation}
where
$2\beta_0 <<1$ is the amplitude, $\omega_0$ is the frequency and $\psi_0$ is the phase of the incoming gravitational wave. In the linear approximation we have to put $L \simeq 1$,
and the variations of the magnetic and electric fields can be rewritten as follows.

\subsubsection{The case $n^2 \equiv 1$}

\begin{equation}
 \delta {\cal B}^1(u) = 0\,, \quad \delta {\cal E}^1(u) =   2\beta(u) B^{(1)} \sin{2\Theta} \,,
 \label{Dphys7}
\end{equation}
\begin{equation}
\delta {\cal B}^2  =  \frac12 B_{\bot} \cos{\Theta} \left\{ \beta(u)\left[2-\cos{2\Theta} + 4 \frac{{B^{(1)}}^2}{B^2_{\bot}} + 4 \frac{\Psi^2_0 {\cal H}_0}{B^2_{\bot}} \right] + v \beta^{\prime}(u) \cos{2\Theta} \right\}
 \,,
 \label{Dphys1}
\end{equation}
\begin{equation}
\delta {\cal B}^3  = - \frac12 B_{\bot} \sin{\Theta} \left\{ \beta(u)\left[2+\cos{2\Theta} + 4 \frac{{B^{(1)}}^2}{B^2_{\bot}} + 4 \frac{\Psi^2_0 {\cal H}_0}{B^2_{\bot}}   \right] - v \beta^{\prime}(u) \cos{2\Theta} \right\}
 \,,
 \label{Dphys2}
\end{equation}
\begin{equation}
 \delta {\cal E}^2  =
 -\frac12 B_{\bot} \sin{\Theta} \left\{ \beta(u)\left[-\cos{2\Theta} + 4 \frac{{B^{(1)}}^2}{B^2_{\bot}} + 4 \frac{\Psi^2_0 {\cal H}_0}{B^2_{\bot}}   \right] - v \beta^{\prime}(u) \cos{2\Theta} \right\}
 \,,
 \label{Dphys3}
\end{equation}
\begin{equation}
 \delta {\cal E}^3 = -\frac12 B_{\bot} \cos{\Theta} \left\{ \beta(u)\left[\cos{2\Theta} + 4 \frac{{B^{(1)}}^2}{B^2_{\bot}} + 4 \frac{\Psi^2_0 {\cal H}_0}{B^2_{\bot}}   \right] + v \beta^{\prime}(u) \cos{2\Theta} \right\}
  \,.
 \label{Dphys4}
\end{equation}
Here we use the constant
\begin{equation}
 {\cal H}_0  \equiv
 {\cal H}(\phi(0)) =  m^2_{({\rm A})} + \lambda \left[3\phi^2(0) - \phi^2_* \right]\,.
 \label{Dphys5}
\end{equation}
Clearly, the variations of the magnetic and electric fields have the frequency $\omega_0$, coinciding with the frequency of the gravitational wave, and have the amplitude proportional to the value of the transversal part of the initial magnetic field.
There are contributions in these variations, which attract a special interest; first of all, we mean the terms in (\ref{Dphys1})-(\ref{Dphys4}) proportional to the function
\begin{equation}
v \beta^{\prime}(u) = - v \beta_0 \frac{\sqrt2}{c} \omega_0 \sin{\left(\frac{\sqrt2}{c} \omega_0 u {+}\psi_0\right)}
 \,,
 \label{Dphys11}
\end{equation}
which describe pure gravitational-wave effect in the electromagnetic field variations. When the coordinate $x^1$ is fixed, the amplitude of this function $ t \beta_0 \omega_0$ grows with time, thus providing resonant-type effect of amplification of the electromagnetic field variations induced by the gravitational wave field.
Another important features relate to the terms in (\ref{Dphys1})-(\ref{Dphys4}), which describe the axionic effects; we mean the terms containing $\frac{\Psi^2_0 {\cal H}_0}{B^2_{\bot}}$.

The GW-induced variation of the pseudoscalar field $\delta \phi$ is linear in the GW-amplitude $\beta$
\begin{equation}
\delta \phi = -2\beta \sin{2\Theta}
 \,.
 \label{varphi1}
\end{equation}
This allows us to estimate the GW-induced variation of the energy of the axion field, $\delta W_{({\rm A})}$ as follows. When $\phi(0){=}0$, using the formula (\ref{TAX}) and non-perturbed value of the velocity, we obtain
$$
W_{({\rm A})}(u) \equiv U^i T_{ik}^{({\rm A})}U^k = \frac12 \Psi^2_0 \left[{\phi^{\prime}}^2(u) + m^2_{({\rm A})}\left(\phi^2(u)-\phi^2_* \right) + \frac12 \lambda \left(\phi^2(u)-\phi^2_* \right)^2 \right] \,,
$$
\begin{equation}
W_{({\rm A})}(0) = \frac12 \Psi^2_0 \left[- m^2_{({\rm A})}\phi^2_*  + \frac12 \lambda \phi^4_* \right] \,,
 \label{varphi3}
\end{equation}
and thus
\begin{equation}
\delta W_{({\rm A})} \equiv W_{({\rm A})}(u)-W_{({\rm A})}(0)= 2 \Psi^2_0 \sin^2{2\Theta} \left[{\beta^{\prime}}^2(u)+
\left(m^2_{({\rm A})} - \lambda \phi^2_* \right) \beta^2 \right]
 \,.
 \label{varphi4}
\end{equation}
For the very illustrative special case $\lambda {=}\frac{2 m^2_{({\rm A})}}{\phi^2_*}$ we obtain that $W_{({\rm A})}(0) {=}0$ and
\begin{equation}
\delta W_{({\rm A})} = 2 \Psi^2_0 \sin^2{2\Theta} \left[{\beta^{\prime}}^2(u)-
m^2_{({\rm A})} \beta^2 \right]
 \,.
 \label{varphi49}
\end{equation}
Using (\ref{Dphys17}) we can present the energy variation averaged over the GW period
as follows
\begin{equation}
\left<\delta W_{({\rm A})} \right> =  \Psi^2_0 \beta^2_0 \sin^2{2\Theta} \left[2 \ \frac{\omega^2}{c^2}-
m^2_{({\rm A})}\right]
 \,.
 \label{varphi41}
\end{equation}
As it will be shown below this quantity is estimated to be negative.

\subsubsection{The case $n^2 \neq 1$}

For this model the weak variations of the magnetic and electric fields can be written in the following form:
\begin{equation}
 \delta {\cal B}^1(u) = 0\,, \quad \delta {\cal E}^1(u) =  - \frac{2}{n^2}\beta(u) B^{(1)} {\cal H}^*_0 \sin{2\Theta}  \,,
 \label{DDphys7}
\end{equation}
\begin{equation}
\delta {\cal B}^2  =  \beta(u) B^{(2)} - \delta {\cal E}^3
 \,,
 \quad
\delta {\cal B}^3  = - \beta(u) B^{(3)} + \delta {\cal E}^2
 \,,
 \label{DDphys2}
\end{equation}
\begin{equation}
 \delta {\cal E}^2  =
 - \frac{2\beta(u)}{(n^2{-}1)} B_{\bot} \sin{\Theta} \left[1 + 2 H^*_0 \cos^2{\Theta}   \right]
 \,,
 \quad
 \delta {\cal E}^3 = - \frac{2\beta(u)}{(n^2{-}1)} B_{\bot} \cos{\Theta} \left[1 + 2 H^*_0 \sin^2{\Theta}   \right]
  \,,
 \label{DDphys4}
\end{equation}
\begin{equation}
 {\cal H}^*_0  \equiv
 \frac{1}{\left[1 + \frac{(n^2-1)}{\mu} \left(\Psi^2_0 m^2_{({\rm A})} + \frac{{B^{(1)}}^2}{\varepsilon B^2_{\bot}}\right)\right]} \,,
 \quad
\delta \phi = \frac{2\beta(u)}{\mu}{\cal H}^*_0 \sin{2\Theta}
\,.
 \label{DDphys51}
\end{equation}
Again we deal with variations of the electric and magnetic fields with the frequency of the gravitational wave. These electromagnetic field variations can be amplified anomalously, when
$|n^2{-}1| \equiv |\chi| <<1$.

\subsection{Possible applications to experiments with natural magnetic fields}

Let us consider a typical application of the model to a physical configuration with stationary magnetic field. For instance, it
could be the interstellar magnetic field, and (in some approximation) the terrestrial magnetic field at
large altitudes. Taking into account formulas
(\ref{Dphys7})-(\ref{DDphys51},) we can assume that periodic
gravitational radiation from, e.g., relativistic binary  ${\rm
J0737{-}3039(A,B)}$ with orbital period ${\rm P}_b {=}
0.102251563$ days and orbital period derivative $\dot{{\rm P}}_b =
{-} 1.21 \cdot 10^{{-}12}$ (see, e.g., \cite{25a,25b,Will}) produces
periodic variations of the magnetic and electric fields with the
frequency $\nu_{{\rm gw}} {=} 2 \nu_b {=} \frac{2}{{\rm P}_b} \simeq 2.3
\cdot 10^{-4} {\rm Hz}$. These variations are modulated by the
Earth rotation. The amplitude of the gravitational waves from this
binary can be estimated as $2\beta_0 \simeq 10^{{-}23}$, thus the
fractional amplitudes of variations can be estimated as
\begin{equation}
\left|\frac{\delta {\cal B}}{B_{\bot}} \right| = 2\beta_0 {\cal Q} =  \left|\frac{\delta {\cal E}}{B_{\bot}} \right|
\,,
\label{Q1}
\end{equation}
where ${\cal Q}$ is the so-called quality-factor, which depends on
the angle $\Theta$. It is convenient to estimate this factor for
angle $\Theta \neq \{0, \frac{\pi}{4}, \frac{\pi}{2}, \pi\}$; for
the illustration we use,  e.g, $\Theta {=} \frac{\pi}{8}$.

\subsubsection{Estimation of the effect magnitude at $n^2 \equiv 1$}

For the model with $n^2 \equiv 1$ and $B^{(1)}{=}0$ this coefficient can be estimated as
\begin{equation}
{\cal Q} \simeq
0.3 +  0.9 \frac{\Psi^2_0 m^2_{({\rm A})}}{B^2_{\bot}} + 0.16 \omega_0 T_0
\,,
\label{Q2}
\end{equation}
where $T_0$ is the time of observation. During one year of
observation ($T_0{=}365$ days) the third term in (\ref{Q2}) could
reach the value $0.7 \cdot 10^{4}$. Let us estimate the second
term. Let us mention, first, that the multiplier $\Psi^2_0$ has
the dimensionality of energy per length; as for the quantity
$m_{({\rm A})}$, it has the dimensionality of inverse length and
is connected with the real mass of axion $m_{({\rm axion})}$ by
the relationship $m_{({\rm A})} {=} \frac{c}{\hbar} m_{({\rm
axion})}$. The parameter $\Psi_0$ is reciprocal to the
axion-photon-photon coupling constant $g_{{\rm A} \gamma \gamma}$,
i.e., $\frac{1}{\Psi_0}{=} g_{{\rm A} \gamma \gamma}$, and
$g_{{\rm A} \gamma \gamma}$ itself is
estimated to belong to the range $10^{-12} {\rm
GeV}^{-1} < g_{{\rm A} \gamma \gamma} < 10^{-5}{\rm
GeV}^{-1}$. For illustration we take the value  $g_{{\rm A} \gamma \gamma} \simeq 10^{-6} {\rm
GeV}^{-1}$. Thus, when we deal with the natural units ($c{=}\hbar
{=}1$) we obtain the value $\Psi_0 \simeq 10^{15} {\rm eV}$;
equivalently in the Gaussian system of units we have $\Psi_0
\simeq 3 \cdot 10^{11} \left[{\rm g}^{\frac12} \cdot {\rm
cm}^{\frac12} \cdot {\rm s}^{-1}\right]$ (for conversion factors
see, e.g., \cite{units}). Keeping in mind the restriction
$10^{-6} {\rm eV} < m_{({\rm A})} < 1 {\rm eV}$, we take for illustration the
value $m_{({\rm axion})}{=}10^{-12} m_{({\rm e})}$ for the mass of
axion. Thus, using the Gaussian system of units we obtain that
$m_{({\rm A})} \to \frac{c}{\hbar} \cdot 10^{-12} m_{({\rm e})}
\simeq 2.7 \cdot 10^{-2} {\rm cm}^{-1}$. Thus the term $\Psi_0 m_{({\rm A})}$ is
of the order of $8 \cdot 10^{9}\left[{\rm g}^{\frac12} \cdot {\rm
cm}^{-\frac12} \cdot {\rm s}^{-1}\right]$. As for the averaged value of the
axion field energy-density  variation (\ref{varphi41}), we obtain that for the GW frequencies of the infra-low range $10^{-3}-10^{-7}$, $\frac{\sqrt2 \omega_0}{c}\simeq 27 \cdot (10^{-14}-10^{-18}){\rm cm}^{-1}$, i.e., the quantity $\left<\delta W_{({\rm A})} \right>$ is negative.

When we deal with the terrestrial magnetic field with $B_{\bot} {=} 0.5 {\rm Gs}$, the second term in (\ref{Q2}) is of the order
$0.9 \frac{\Psi^2_0 m^2_{({\rm A})}}{B^2_{\bot}} \simeq 2 \cdot 10^{20}$.
This contribution describes the leading order term in ${\cal Q}$, and one should stress that it relates to
the contribution connected with the axion field.

To conclude, we have to say, that the gravitational radiation from
the binary system  ${\rm J0737}$ - ${\rm 3039(A,B)}$ provides the
periodic variations of the magnetic field and the appearance of electric field variations
with the frequency
$\nu_{{\rm gw}} \simeq 2.3 \cdot 10^{-4} {\rm Hz}$. The appearance of the axion induced electric field is a new result of model analysis;
the magnitude of this electric field is estimated
to be of the order $\delta {\cal E} \simeq  10^{-3} \ \frac{ {\rm statV}}{{\rm cm}}$ for the
terrestrial magnetic field.

\subsubsection{Estimation of the effect magnitude at $n^2 \neq 1$}

For the case $n^2 \neq 1$ one obtains that ${\cal Q}
{=}\frac{1}{(n^2{-}1)}$. When we deal with vacuum in the standard
sense, i.e., the gas and charged particles are removed from the
device, the quantity $n^2{-}1 \equiv \chi$ is predetermined by the
dark matter susceptibility only. In classical molecular physics
the quantity $\chi$ is presented by the formula $\chi {=}
\frac{4\pi}{3} \alpha N$, where $N$ is the molecule number per
unit volume, and $\alpha$ is the susceptibility of an individual
molecule. For axionic dark matter in the Earth environment one can
assume that the axion mass-density is $\rho_{({\rm DM})} \simeq
0.033 \ M_{({\rm Sun})} {\rm pc}^{-3}$, or in the natural units
$\rho_{({\rm DM})} \simeq 1.25 \ {\rm GeV} \cdot {\rm cm}^{-3}$.
Then the axion number density is estimated to be $N_{({\rm A})}
\simeq 10^{15}  {\rm cm}^{-3}$. The axionic susceptibility
$\alpha_{({\rm A})}$ is much less than the one for molecules,
$\alpha_{({\rm molecule})} \simeq 10^{-24} {\rm cm}^3$, thus
$\chi_{({\rm A})}<< 10^{-9}$. This means that the corresponding
quality-factor is characterized by the value ${\cal Q} >>
10^{9}$, and we prefer to estimate  this factor as ${\cal Q}
\simeq 10^{20}$ keeping in mind that electromagnetic interactions
are $10^{11}$ times stronger than weak interactions. Thus,
gravitational waves from the binary system ${\rm
J0737{-}3039(A,B)}$  can produce periodic variations of the electric
field, which have the amplitude of the order $\left|\delta {\cal
E}\right| \simeq 10^{-3} \ \frac{{\rm statvolt}}{{\rm cm}}$ for the
terrestrial magnetic field.

\subsubsection{Constraints of the model}

We used three assumptions, which restrict estimations of the
magnitude of the predicted effect. Let us discuss them shortly.

\vspace{2mm}
\noindent
{\it (i)} Since we predict an anomalous growth of the electric field provoked
by the axion-photon interactions under the influence of a periodic gravitational
wave, we should estimate whether the generated electric and magnetic field variations
on the GW frequency violate the first model assumption that the electromagnetic field is the test one.
In other words, is it necessary to modify the gravity field equations (\ref{GR}) in order to take into account
the {\it feedback} of the generated electric and magnetic field? Clearly, we have to compare
the maximum value of the contribution of the weak GW - field into the Einstein tensor (left-hand side of (\ref{GR})) and
the maximum value of the contribution of the generated electric and magnetic fields into the total stress-energy tensor (right-hand side of in (\ref{GR})).
This estimation uses the stress-energy tensor $T^{({\rm EM})}_{ik}$ (\ref{u8}), and it can be written as follows
\begin{equation}
\beta_0 \left(\frac{2\pi\nu_{{\rm gw}}}{c}\right)^2 >> \frac{8\pi G}{c^4}|B_{\bot}||\delta {\cal B}| \ \ \rightarrow \ \ 2\beta_0 \frac{8\pi G}{c^4}|B_{\bot}|^2 {\cal Q}
\,.
\label{rest1}
\end{equation}
Using the optimistic estimation ${\cal Q}{=}10^{20}$, and the  value  $|B_{\bot}|{=} 0.5 {\rm Gs}$ for the terrestrial magnetic field we obtain that
\begin{equation}
\nu^2_{{\rm gw}} >> \frac{4 G}{\pi c^2}|B_{\bot}|^2 {\cal Q} \ \ \rightarrow \ \ 0.25 \cdot 10^{-8} \left\{\frac{|B_{\bot}|}{0.5 {\rm Gs}} \right\}^2 [s^{-2}]
\,.
\label{rest2}
\end{equation}
Since for the relativistic binary  ${\rm
J0737{-}3039(A,B)}$ we have $\nu^2_{{\rm gw}} \simeq 5.3 \cdot 10^{-8} {\rm Hz}^2$ we can confirm that we are still working in the range of validity of the model, when consider the example of terrestrial magnetic field. The interstellar magnetic field is estimated to be of the order $10^{-6}{-}10^{-4} {\rm Gs}$, thus, this model is also valid for the description of critical phenomena in the magnetized interstellar medium.

\vspace{2mm}
\noindent
{\it (ii)} We considered the ideal model of electrodynamic system without spatial boundaries. This assumption is valid, when the size of the electrodynamic system, say ${\cal R}$, is much bigger that the wave-length of the gravitational radiation, $\lambda_{{\rm gw}}$, i.e., ${\cal R}>> \lambda_{{\rm gw}}$. This model seems to be appropriate for cosmic plasma or magnetized cosmic medium through which the gravitation wave emitted by astrophysical sources are traveling. In the realistic models associated with terrestrial devices or with geo-magnetic field the size of electrodynamic system is finite and
we have to solve the appropriate boundary value problem. In this case the variations of electric, magnetic and axion fields should be the functions of $u$, $v$, $x^2$ and $x^3$.
This means that the stationary solution depending on the retarded time $u$, which we discussed above, could be reached only asymptotically (when the contributions of the boundary regime vanish because of damping). We hope to consider the boundary-value problem and to estimate the time necessary for the transition to the stationary regime in a special work. Nevertheless, it is clear, that  this time depends on the period and amplitude of the GW- field, and is rather big. That is why, choosing the appropriate illustration, we focus on the terrestrial magnetic field, which has the modern structure at least during ten millenniums, and the stationary GW-source, which certainly exists during this period. Magnets, which were switched on not long ago, seem to be not too prospective in this sense.

\vspace{2mm}
\noindent
{\it (iii)}
We considered the initial magnetic field to be homogeneous. This assumption seems to be valid for magnetized interstellar medium, but it is an idealization, when we deal with dipole-type terrestrial magnetic field. Nevertheless, the analysis of such homogeneous model is a good first step to find out the possibility of anomalous behavior of electrodynamic system in the axionic environment.

\section{Discussion and conclusions}

\noindent {\it 1. On the new mechanism of axion-photon-graviton coupling}

\noindent
We described one possible mechanism of an anomalous response of an electromagnetic field on the action of a gravitational wave, in which the axionic dark matter plays
a role of mediator-amplifier. The mechanism works as follows. Let the pure magnetic field be static, when the gravitational wave is absent. Since there is no electric field is such system, the pseudo-invariant $I^{*} \equiv \frac14 F^{*}_{ik}F^{ik} {=} \frac12 B_m E^m$ is equal to zero, and the pseudoscalar (axion) field $\phi$ is unperturbed being equal to constant $\phi(0)$. In this case the axion field is a hidden field from the point of view of axion electrodynamics. When the periodic gravitational wave appears, it deforms the initially static magnetic field and thus generates an electric field, which was absent in the static situation. Then the pseudo-invariant $I^{*}$, being non-vanishing now, becomes the periodic source for the axion field, thus providing backreaction of the axion field on the electromagnetic one via the standard mechanism of the axion-photon coupling.
As a result, in addition to variations of the magnetic field, variations of electric field appear, being periodic functions of time with the frequency coinciding with the frequency of the incoming gravitational wave. Measurements of these electric field variations on the well-known frequency could be the base of strategy of new experiments, in which one could verify both hypotheses about gravitational wave and axionic dark matter existence.

\vspace{2mm}
\noindent {\it 2. On the symmetry of the effect}

\noindent
The magnitude of the electromagnetic response depends essentially on the angles between the direction of the initial magnetic field, on the one hand, and the gravitational wave-vector, as well as, the gravitational wave polarization eigen-vectors, on the other hand. For instance, when the initial magnetic field is orthogonal to the gravitational wave front plane (i.e., $B_{\bot}{=}0$), the described effect is absent; because of the Earth rotation one can find such gravitational wave sources, for which every 24 hours the magnetic field direction coincides with the direction to the gravitational source, so that the described effect vanishes thus providing the existence of the so-called null-point readout for the electric field detector.

\vspace{2mm}
\noindent {\it 3. On the magnitude of the effect}

\noindent
The described mechanism of axion-photon-graviton coupling can be characterized as anomalous, and we have to mention three details in this connection.

First, we can use the term anomalous (or critical), since we obtained exact solutions
(\ref{2phys1})-(\ref{2phys4}), which display the following behavior.
If we take the key multiplier $\left\{\frac{\sinh{\beta}}{(n^2{-}1)}\right\}$ and calculate the limit of
$\beta \to 0$ and then $n^2\to 1$ we obtain that this double limit
is equal to zero, $\lim_{n^2 \to 1} \lim_{\beta \to
0}\left\{\frac{\sinh{\beta}}{(n^2{-}1)}\right\} {=}0$. If we take
first the limit of $n^2\to 1$ and then $\beta \to 0$, the double
limit $ \lim_{\beta \to 0} \lim_{n^2 \to 1}
\left\{\frac{\sinh{\beta}}{(n^2{-}1)}\right\} {=}\infty$ gives infinity.
Since these two double limits do not coincide, we can speak of a
critical behavior of the electromagnetic field near the singular
point $n^2{=}1$; when $n^2$ tends to one, the response grows anomalously. This effect was
also predicted in \cite{BL} for the case of axion absence; the novelty of the model under
discussion is that the susceptibility
$\chi = n^2{-}1$ is now prescribed to the axionic dark matter, and this assumption gives us
renewed estimation of the effect, which could be now of the order of
$\left|\frac{\delta {\cal E}}{B_{\bot}} \right| \simeq 10^{-3}$.

Second, in case when $n^2 \equiv 1$, the obtained exact solutions
(\ref{phys2})-(\ref{phys5}) contain  the terms linear in time. In
other words, the electric response grows with time, and can give
the amplification coefficient ${\cal Q} \simeq 0.7 \cdot 10^{4}$
for an one year permanent monitoring. This effect is inherited
from the theory, in which axions are absent \cite{BL}.

Third, a principally new contribution into the electromagnetic
response appeared due to the axion-photon interaction in the field
of gravitational radiation is described in
(\ref{phys2})-(\ref{phys6}) by the terms containing
$\frac{\Psi^2_0 m^2_{({\rm A})}}{B^2_{\bot}}$. When $\Psi^2_0
m^2_{({\rm A})} \neq 0$, $\beta(u) \neq 0$ and $B_{\bot} \neq 0$,
the function $Z(u)$ (see the formula for the electromagnetic
response (\ref{phys2})-(\ref{phys6}) ) contains $B^2_{\bot} \neq
0$ in the denominator. This means that at presence of the
gravitational wave the electromagnetic response grows anomalously,
when $B^2_{\bot} \to 0$. Nevertheless, when $B^2_{\bot} \equiv 0$
identically, the effect vanishes. Again, we deal with critical
behavior of the response, since $\lim_{B_{\bot}\to
0}\{F_{ik}(B_{\bot}) \} {=} \infty \neq \{F_{ik}(B_{\bot}{=}0)
\}$, when $\beta \neq 0$. Optimistic values for the term
$\frac{\Psi^2_0 m^2_{({\rm A})}}{B^2_{\bot}}$ are estimated to be
of the order $10^{20}$ for the terrestrial magnetic field, and of the order $10^{28}$ for the magnetized
interstellar medium.

Let us mention that the described critical behavior can be interpreted in terms of phase transition of the second kind (see our reasoning presented in Section 5.4).

\vspace{2mm}
\noindent{\it 4. On the energy balance in the light of anomalous growth of the electric field}

\noindent
The natural question arises: what is the energy reservoir for the support of anomalous growth of the electric
field described above? In principle, there are three possible candidates: first, the external gravitational-wave field;
second, the pseudoscalar (axion) field; third, the recoil of the material medium, in which the electromagnetic field is distributed.
Our opinion is that the main source of the anomalous behavior is the energy reservoir granted by the axionic dark matter, described
in this model by the pseudoscalar field. Our explanation is based on the formula (\ref{varphi41}), which shows that the variation
of the energy-density of the pseudoscalar field is negative and rather big, since the parameter $\Psi^2_0$ is estimated to be large.
Thus, just the axionic dark matter supplies the growth of the electromagnetic field via the described above mechanism of axion-photon coupling.
The gravitational field in this mechanism plays the role of catalyzer, which provides the non-static background.

\vspace{2mm}
\noindent {\it 5. Prospects}

\noindent
We presented exact solutions to the system of equations of axion electrodynamics. These solutions describe some stationary regime established after the
gravitational wave appearance; this regime is described by the functions depending on the retarded time only, $\phi(u)$. We understand that in order to
convince colleagues that such anomalous behavior is possible, in the nearest future we have to elaborate in detail the model of transition from a static
state $\phi {=}\phi(0)$ to the stationary state $\phi {=}\phi(u)$ and to estimate the typical time of this transition. Also we made some idealistic
estimations of the new effect of axion-photon-graviton coupling. Our goal was to attract attention to this results, and we, of course, understand
that in nearest future we should construct more realistic models for the GW-induced electromagnetic response of the terrestrial magnetic
field and magnetized interstellar medium. These tasks are very inspirational, and studying them we keep in mind two unsolved physical problems. First of all,
detection of predicted electric signal generated by magnetic field (interstellar or terrestrial) in an axionic environment might be an indirect proof of
axionic dark matter existence; quantitative characteristics of detected electric signal would give the constraints for the axion mass for the constant of
axion-photon coupling. Second, if the detected electric signal generated by magnetic field (interstellar or terrestrial) would have the specific gravitational-wave frequency,
we would obtain new (indirect) arguments for the gravitational wave existence. Let us stress again, that the predicted electromagnetic signal can be indicated as anomalous, and thus
its detection can attract the attention of experimentalists.

\vspace{3mm}

\noindent
{\bf Acknowledgments}

\noindent We thank the National Natural Science Foundation of
China (Grant No. 10875171) and Russian Foundation for Basic
Research (Grant No. 14-02-00598) for support.
This work was initiated in 2009 during the visit of A.B. to the
Center for Gravitation and Cosmology of Purple Mountain
Observatory of Chinese Academy of Science, Nanjing; A.B. is
grateful to colleagues from this Center for invitation and
hospitality. W-T.N. also thanks the National Science Council of
the Republic of China for support (Grant No. NSC102-2112-M-007-019).


\begin{thebibliography}{50}

\bibitem{PVLAS} Zavattini E et al 2008 New PVLAS results and limits on magnetically induced optical rotation and ellipticity in vacuum Phys. Rev. D 77 032006

\bibitem{PVLAS2} Zavattini G et al 2012 Measuring the magnetic birefringence of vacuum: the PVLAS experiment Int. J. Mod. Phys. A 28 1260017

\bibitem{GammeV} Chou A S et al (GammeV Collaboration) 2008 Search for axionlike particles using a variable-baseline photon-regeneration technique Phys. Rev. Lett. 100 080402

\bibitem{CAST} Andriamonje S et al (CAST Collaboration) 2007 An improved limit on the axion–photon coupling from the CAST experiment JCAP 0704, 010

\bibitem{CAST2} Arik, M et al (CAST Collaboration) 2011 Search for sub-eV mass solar axions by the CERN axion solar telescope with He-3 buffer gas Phys. Rev. Lett. 107 261302

\bibitem{OSQAR} Pugnat P et al (OSQAR Collaboration) 2008 Phys. Rev. D 78, 092003

\bibitem{QA}  Chen S-J, Mei H-H and Ni W-T 2007 Q \& A experiment to search for vacuum dichroism, pseudoscalar-photon interaction and millicharged fermions Mod. Phys. Lett. A 22 2815 (arXiv:hep-ex/0611050)

\bibitem{QA2} Ni W-T, Mei H-H and Wu S-J 2013 Foundations of classical electrodynamics, equivalence principle and cosmic interactions: a short exposition and an update Mod. Phys. Lett. A 28 1340013 (arXiv:gr-qc/1204.0872)

\bibitem{BMV} Battesti R et al 2008 The BMV experiment: a novel apparatus to study the propagation of light in a transverse magnetic field Eur. Phys. J. D 46 323 (arXiv:0710.1703)

\bibitem{BMV2} Battesti R et al 2010 Photon regeneration experiment for axion search using X-Rays Phys. Rev. Lett. 105 250405

\bibitem{PQ} Peccei R D and Quinn H R 1977 CP conservation in the presence of pseudoparticles Phys. Rev. Lett. 38 1440

\bibitem{Weinberg} Weinberg S 1978 A new light boson? Phys. Rev. Lett. 40 223

\bibitem{Wilczek} Wilczek F 1978 Problem of strong P and T invariance in the presence of instantons Phys. Rev. Lett. 40 279

\bibitem{ADM1} Raffelt G G 1990 Astrophysical methods to constrain axions and other novel particle phenomena
Phys. Rep. 198 1

\bibitem{ADM2} Turner M S 1990 Windows on the axion  Phys. Rep. 197 67

\bibitem{ADM3}  Battesti R  et al 2008 Axion searches in the past, at present, and in the near future
Lect. Notes Phys. 741 199 (arXiv:0705.0615)

\bibitem{Ni} Ni W-T 1977 Equivalence principles and electromagnetism Phys. Rev. Lett. 38 301

\bibitem{Sikivie} Sikivie P 1983 Experimental Tests of the "Invisible" Axion Phys. Rev. Lett. 51 1415

\bibitem{Wlczk} Wilczek F 1987 Two applications of axion electrodynamics Phys. Rev. Lett.  58 1799

\bibitem{Ni2} Ni W-T 2008 From equivalence principles to cosmology: cosmic
polarization rotation, CMB observation, neutrino number asymmetry,
Lorentz invariance and CPT Prog. Theor. Phys. Suppl. 172 49  (arXiv:0712.4082)

\bibitem{boc}  Bocaletti D, De Sabbata V, Fortini P and Gualdi C 1970 Conversion of photons
into gravitons and vice versa in a static electromagnetic field
Nuovo Cimento B 70  129

\bibitem{zel1}  Zel'dovich Ya B 1974  Electromagnetic and gravitational waves in a stationary magnetic field Sov. Phys. JETP 38 652

\bibitem{BV} Balakin A B and Vakhrushev D V 1993  Critical character of gravitational wave modulation of electric and magnetic fields in isotopic media Russian Physics Journal 36  833

\bibitem{BL} Balakin A B and  Lemos J P S 2001 Singular behaviour of electric and magnetic fields in dielectric media in a nonlinear gravitational wave background Class. Quantum Grav. 18  941

\bibitem{CQG2007} Balakin A B 2007 Magnetic relaxation in the Bianchi-I universe
Class. Quantum Grav. 24 5221

\bibitem{GC2007} Balakin A B 2007  Extended Einstein-Maxwell model Gravitation and Cosmology 13 163

\bibitem{BMZ} Balakin A B, Muharlyamov R K and Zayats A E 2014 Nonminimal Einstein-Maxwell-Vlasov-axion model Class. Quantum Grav. 31 025005

\bibitem{tds} Brown J D 1993 Action functionals for relativistic perfect fluids Class. Quantum Grav. 10 1579

\bibitem{MTW}
Misner C W, Thorne K S  and  Wheeler J A 1973 {\it Gravitation} (San Francisco: Freeman)

\bibitem{ExactSolutions} Stephani H, Kramer D, MacCallum M A H,
Hoenselaers C and Herlt E 2003 {\it Exact Solutions to Einstein's Field
Equations} (New York: Cambridge University Press)

\bibitem{BN10} Balakin A B and Ni W-T 2010 Non-minimal coupling of photons and axions Class. Quantum Grav. 27 055003

\bibitem{Goursat}
Goursat E A 1923 {\it Course in Mathematical Analysis III: Variation of Solutions and Partial Differential Equations of the Second Order and Integral Equations and Calculus of Variations} (Paris: Gauthier-Villars)

\bibitem{Amaldi1} Balakin A B 2000 Evolution of relativistic hierarchical systems in the field of gravitational radiation Annalen der Physik (Special Issue) 9 21

\bibitem{Amaldi2} Balakin A B 1995 Hierarchical approach to the theory of detection of periodic gravitational radiation
{\it Proc. Ist Edoardo Amaldi Conference on Gravitational Wave Experiments} 269 (Singapore: World Scientific)

\bibitem{LL5} Landau L D and Lifshitz E M 1980 {\it Statistical Physics, V} {Oxford: Butterworth-Heinemann}

\bibitem{25a} Burgay M. et al 2003  An increased estimate of the merger rate of double neutron stars from observations of a highly relativistic system
 Nature  426 531

\bibitem{25b} Lyne A G et al 2004 A double-pulsar system - a rare laboratory for relativistic gravity and plasma physics Science 303 1153

\bibitem{Will} Will C 2005 The Confrontation between General Relativity and Experiment. Living Rev. Rel. 9 3

\bibitem{units} Dominguez-Tenreiro R and Quir\'os  M 1988 {\it An introduction to cosmology and particle physics}(Singapore: World Scientific Publisging)


\end{thebibliography}
\end{document}